\documentclass[11pt]{article}
\pdfoutput=1
\usepackage{amsmath}
\usepackage{amsfonts}
\usepackage{amssymb}
\usepackage{mathptmx}
\usepackage{slashed}
\usepackage{latexsym}
\usepackage{graphicx}
\usepackage{color}
\usepackage{caption}
\usepackage{subcaption}
\usepackage{multicol}

\newcommand{\newc}{\newcommand}
\newc{\be}{\begin{equation}}
\newc{\ee}{\end{equation}}
\newc{\bea}{\begin{eqnarray}}
\newc{\eea}{\end{eqnarray}}
\newc{\ol}{\overline}
\newc{\wt}{\widetilde}
\newc{\bs}{\boldsymbol}
\newc{\m}{\mathcal}
\newc{\ra}{\rightarrow}
\newc{\lra}{\leftrightarrow}
\newc{\ba}{\begin{eqnarray}}
\newc{\ea}{\end{eqnarray}}
\newc{\pa}{\partial}
\newc{\D}{\Delta}

\newc{\nn}{\nonumber}

\newc{\La}{\mathcal{L}_1}
\newc{\Lac}{\mathcal{L}_1^{-1}}
\newc{\Lb}{\mathcal{L}_2}
\newc{\Lbc}{\mathcal{L}_2^{-1}}
\newc{\Lc}{\mathcal{L}_3}
\newc{\Lcc}{\mathcal{L}_3^{-1}}
\newc{\ot}{\otimes}

\def\beq{\begin{equation}}
\def\eeq{\end{equation}}
\def\bea{\begin{eqnarray}}
\def\eea{\end{eqnarray}}

\newcommand{\cc}{\sp*}

\newcommand{\pr}{\sp\prime}
\newcommand{\ov}{\overline}

\begin{document}

\begin{titlepage}

%\vspace*{-15mm}
%\begin{flushright}
%SHEP-11-XX\\
%\end{flushright}
\vspace*{0.7cm}

\begin{center}
{
\bf\LARGE
Gauge Coupling Unification in $E_6$ F-Theory GUTs 
\\
\vspace{0.05in}
with Matter and Bulk Exotics from Flux Breaking}
\\[12mm]
James~C.~Callaghan$^{\star}$
\footnote{E-mail: \texttt{James.Callaghan@soton.ac.uk}},
Stephen~F.~King$^{\star}$
\footnote{E-mail: \texttt{king@soton.ac.uk}},
George~K.~Leontaris$^{\dagger}$
\footnote{E-mail: \texttt{leonta@uoi.gr}},
\\[-2mm]

\end{center}
\vspace*{0.50cm}
\centerline{$^{\star}$ \it
School of Physics and Astronomy, University of Southampton,}
\centerline{\it
University Road, SO17 1BJ Southampton, United Kingdom }
\vspace*{0.2cm}
\centerline{$^{\dagger}$ \it
Physics Department, Theory Division, Ioannina University,}
\centerline{\it
GR-45110, Ioannina, Greece}
\vspace*{1.20cm}

\begin{abstract}
\noindent We consider gauge coupling unification in $E_6$ F-Theory Grand Unified Theories
(GUTs) where $E_6$ is broken to the 
Standard Model (SM) gauge group using fluxes. In such models there are two types of exotics that can affect gauge coupling unification, namely matter exotics from the matter curves in the 27 dimensional representation of $E_6$ and the bulk exotics from the 
adjoint 78 dimensional representation of $E_6$.
We explore the conditions required for either the complete or partial
removal of bulk exotics from the low energy spectrum. In the latter case 
we shall show that (miraculously) gauge coupling
unification may be possible even if there are bulk exotics
at the TeV scale. Indeed in some cases it is necessary for bulk exotics to survive to the TeV scale in order to 
cancel the effects coming from other TeV scale matter exotics which would by themselves
spoil gauge coupling unification. The combination of matter and bulk exotics in these cases can lead to precise gauge coupling unification which would not be possible with either type of exotics considered by themselves.
The combination of matter and bulk exotics at the TeV scale represents a unique and striking signature of $E_6$ F-theory GUTs that can be tested at the LHC.
 \end{abstract}

 \end{titlepage}

\thispagestyle{empty}
\vfill
\newpage

\setcounter{page}{1}

\section{Introduction}

Recently there has been much interest~\cite{Donagi:2008ca,Beasley:2008dc,Donagi:2008kj,Beasley:2008kw,Blumenhagen:2009yv,Vafa:1996xn} in formulating Grand Unified Theories (GUTs) within the framework of F-theory (for reviews see \cite{Denef:2008wq,Weigand:2010wm,Heckman:2010bq,Grimm:2010ks,Leontaris:2012mh,Maharana:2012tu}).  In this setting, there has been great progress in both global and local model building in the last few years~\cite{Heckman:2008qa}-\cite{Grimm:2011tb}, where global models focus on the construction of elliptically fibered Calabi-Yau four-folds, and local models deal with the effective field theory where the GUT symmetry is realised on a 7-brane wrapping a 4-dimensional surface S.  The so called `semi-local' approach imposes constraints from requiring that S is embedded into a local Calabi-Yau four-fold, which in practice leads to the presence of a local $E_8$ singularity \cite{Heckman:2009mn}.  All Yukawa couplings originate from this single point of $E_8$ enhancement, and we can learn about the matter and couplings of the semi-local theory by decomposing the adjoint of $E_8$ in terms of representations of the GUT group and the perpendicular gauge group.  In terms of the local picture, matter is localised on curves where the GUT brane intersects other 7-branes with extra U(1) symmetries associated to them, with this matter transforming in bi-fundamental representations of the GUT group and the U(1).  Yukawa couplings are then induced at points where three matter curves intersect, corresponding to a further enhancement of the gauge group.

A full classification of how $E_6$, SO(10) and SU(5) GUT groups arise in the semi-local picture has been presented in \cite{Callaghan:2011jj}, where the homology classes of the matter curves were calculated in each case through the spectral cover formalism.   However, as well as matter transforming in the fundamental representation of the GUT group localised on curves on S, in all these cases there will also be bulk matter, coming { from}  the adjoint representation of the GUT group.  In the case that the GUT group is broken down to the Standard Model (SM) gauge group by flux, there are topological formulae which dictate the multiplicities of these adjoint states \cite{Beasley:2008dc}.  It was demonstrated in \cite{Beasley:2008kw} that when the GUT group is SU(5), bulk matter with exotic charges under the SM gauge group can be eliminated from the spectrum provided certain topological properties of the manifold are satisfied.  However, the same study pointed out that when the GUT group is SO(10) or higher, some bulk exotics must always be present in the low energy spectrum.  As such, in order to give these exotics masses, we can look for the topological requirements for them to appear in vector-like pairs, and then turn on VEVs for suitable singlets.  The presence of these bulk states in the spectrum will clearly affect the running of the gauge couplings and their unification, and in \cite{Leontaris:2009wi} it was shown that states descending from the adjoint of SU(5) with exotic SM charges must be completely removed from the spectrum (in the way of \cite{Beasley:2008kw}) due to RGE arguments.

In this paper we will consider models where the GUT group is $E_6$, and is broken by flux breaking down to the Standard Model gauge group via the sequence of breakings
\bea
E_6 & \rightarrow & SO(10)\times U(1)_{\psi} \label{SO(10)} \\
SO(10) & \rightarrow & SU(5)\times  U(1)_{\chi}  \label{SU(5)0} \\
SU(5) & \rightarrow & SU(3) \times SU(2) \times U(1)_Y \label {SM0}.
\eea
In addition we shall consider models where $U(1)_{\psi}$ and $U(1)_{\chi}$ are both broken near the GUT scale
by the vacuum expectation value (VEV) of some scalar field 
or where a particular linear combination,
under which the right handed neutrinos have no charge, survives down to the TeV scale,
namely \cite{King:2005jy,King:2005my},
\begin{equation}
U(1)_N =\frac{1}{4}U(1)_{\chi} +\frac{\sqrt{15}}{4}U(1)_{\psi} \label{U(1)N}.
\end{equation}
%In terms of F-theory model building, the case of successive flux breaking starting with $E_6$ can be studied in the so called spectral %cover formalism \cite{Dudas:2010zb}, where the model building choices relating to the particle content of the model amount to making %choices about the flux breaking. 
Unlike in \cite{Callaghan:2012rv} where the $E_6$ breaking down to $SU(5)$ was assumed to be
achieved by Higgs breaking, and only the last step, namely the $SU(5)$ breaking was due to flux breaking,
here the entire breaking of $E_6$ down to the Standard Model gauge group (perhaps also including a surviving $U(1)_N$) will be achieved in one step by flux breaking. According to the above
discussion, this will necessarily involve bulk exotics appearing below the string scale, which
will be a principal concern 
of the present paper. 

We first focus on the bulk exotics coming from the adjoint 78 dimensional
representation of $E_6$, and look at how topological properties of the internal manifold restrict the elimination of these exotics from the spectrum, and dictate the numbers of exotics which cannot be removed.  
These constraints are then translated into topological restrictions, which then determine the multiplicities of vector-like matter. 
We impose constraints that exotic matter should appear in vector-like pairs and hence can be eliminated from the low energy spectrum by turning on VEVs for appropriate singlet fields.  We show that it is possible that all bulk exotic as well
as matter exotics could have masses close to the GUT scale leading to an MSSM type theory 
somewhat below the GUT scale.
However, there is the possibility that the bulk exotics from 5s of SU(5) could get TeV scale masses whereas those from 10s could be near the GUT scale,
leading to a characteristic spectrum involving TeV vector-like pairs of $d^c$-like and $H_d$-like bulk exotics, with the distinguishing feature that there will always be one more vector pair of $H_d$-like states than $d^c$-like states. 
Although such bulk exotics would by themselves spoil gauge coupling unification, when combined with 
matter exotics, corresponding to having complete 27 dimensional representations of $E_6$ at the TeV scale,
gauge coupling unification is restored. We emphasise that, without such bulk exotics,
the TeV scale matter exotics would lead to an unacceptable splitting of the couplings,
and it is only the combination of TeV scale matter exotics from the 27s plus TeV scale 
bulk exotics from the 78 which (miraculously) restores gauge coupling unification.
The resulting TeV scale matter exotics plus bulk exotics is equivalent 
to four extra $5+\overline{5}$ vector pairs of SU(5),
beyond the minimal supersymmetric standard model (MSSM) spectrum. 
The characteristic prediction of F-theory $E_6$ GUTs of the matter content of 
four extra $5+\overline{5}$
vector pairs can be tested at the LHC.
This may be compared to the equivalent of three extra $5+\overline{5}$ vector pairs predicted by the E6SSM \cite{King:2005jy,King:2005my}.

The layout of the remainder of the paper is as follows.
In Section 2 we review the basic issues related to bulk exotics,
including topological formulae from \cite{Beasley:2008kw}, before applying these ideas to the $E_6$ case, and working out the topological constraints.  These constraints are then translated into relations between the multiplicities of bulk exotics which appear in vector-like pairs. Section 3 is concerned with gauge coupling unification, 
including a renormalisation group equation 
(RGE) analysis, taking into account the constraints on exotics, and the dependences on the exotic masses of the GUT scale and splitting of the gauge couplings are studied.  In Section 4 we discuss $E_6$ models from F-theory,
where the bulk exotics are put into the context of two realistic models given in \cite{Callaghan:2011jj} and \cite{Callaghan:2012rv}. In particular we discuss the possibility that some 
bulk exotics could survive down to the TeV scale, and show how, together with the matter
exotics predicted by these models, they restore gauge coupling unification.

\section{Review of issues related to bulk exotics}
\subsection{Formalism and $SU(5)$ example}
\noindent In F-theory constructions, the appearance of matter is closely related to the topological properties of the internal space.  The multiplicities of states are given by specific topological formulae, and therefore are subject to constraints which have to be taken into account.  Bulk exotic matter arises from the decomposition of the adjoint of the GUT group $G_S$.  When the gauge group $G_S$ is broken to a group $\Gamma_S$ by turning on fluxes in a subgroup $H_S$, with $G_S \supset \Gamma_S \times H_S$, the adjoint of $G_S$ decomposes into representations $(\tau_j , T_j)$ of $\Gamma_S \times H_S$, 

\begin{equation}
\rm{ad}(G_S ) \cong \oplus_{j} \left( \tau_j \otimes T_j \right)
\end{equation}

\noindent Assuming that S is a del Pezzo surface, the multiplicity of four-dimensional massless fields transforming in a representation $\tau_j$ of $\Gamma_S$ is given in terms of the Euler characteristic by

\begin{align}
%\begin{split}
n_j &= - \chi (\mathcal{L}_j , S) % \\&
  =-\left( 1+\frac{1}{2} c_1 (\mathcal{L}_j) \cdot (c_1 (\mathcal{L}_j) +c_1 (S))\right) \label{mult1}
%\end{split}
\end{align}

\noindent where $\mathcal{L}_j$ is a line bundle transforming in the representation $T_j$ of $H_S$, and the topological quantities $c_1 (\mathcal{L}_j)$, $c_1 (S)$ are the first Chern classes of $\mathcal{L}_j$ and S.  The multiplicity of the conjugate representation can be found by noting that $c_1 (\mathcal{L}_j^{-1}) = -c_1 (\mathcal{L}_j)$, leading to the equation

\be
n_j \cc = - \chi (\mathcal{L}_j^{-1} , S) = -\left( 1+\frac{1}{2} c_1 (\mathcal{L}_j) \cdot c_1 (\mathcal{L}_j) - \frac{1}{2} c_1 (\mathcal{L}_j) \cdot c_1 (S) \label{mult2}
\right)\ee

\noindent In the case where we are dealing with states which transform in a representation of $H_S$ corresponding to a direct product of line bundles so that $\mathcal{L}_j = \mathcal{L} \otimes \mathcal{L} \pr$, we have  $n_j   = - \chi (\mathcal{L} \otimes \mathcal{L} \pr , S)$ where
\begin{align}
\chi (\mathcal{L} \otimes \mathcal{L} \pr , S) & = 1+\frac{1}{2} \left\{ c_1 (\mathcal{L}) \cdot c_1 (S) \oplus c_1 (\mathcal{L} \pr) \cdot c_1 (S)\right\} \nn \\
&+\frac{1}{2} \left\{c_1 (\mathcal{L}) \cdot c_1 (\mathcal{L}) \oplus c_1 (\mathcal{L} \pr) \cdot c_1 (\mathcal{L} \pr)\right\} \label{mult2a}
\end{align}
Taking for example the exotics coming from the adjoint of SU(5) after hypercharge flux breaking to the Standard Model, we have the decomposition

\be
24 \rightarrow (8,1)_0 + (1,3)_0 +(3,2)_{-\frac{5}{6}} + (\ov{3},2)_{\frac{5}{6}} \label{SU(5)adj}
\ee

\noindent where the line bundle $\mathcal{L}_Y^{\frac{5}{6}}$ is associated to the hypercharge.  This decomposition gives rise to the states $(3,2)_{-\frac{5}{6}}$ and $(\ov{3},2)_{\frac{5}{6}}$ which are in exotic representations of the SM gauge group.  It has been shown in \cite{Leontaris:2009wi} that the presence of these exotics lower the unification scale to unacceptable values, so we must require that these states are not present in the spectrum.  Using Eqs. (\ref{mult1}) and (\ref{mult2}), and labelling the multiplicities of $(3,2)_{-\frac{5}{6}}$ and $(\ov{3},2)_{\frac{5}{6}}$ states by $m$ and $m \cc$ respectively, we have

\begin{align}
m -m \cc&=- c_1 (\mathcal{L}_Y) \cdot c_1 (S) \label{Ydiff} \\
m+m \cc &=-( 2+c_1 (\mathcal{L}_Y) \cdot c_1 (\mathcal{L}_Y)) \label{Ysum}
\end{align}

\noindent If we require there to be only vector-like pairs of bulk exotics in the spectrum, Eq. (\ref{Ydiff}) tells us that the following dot product has to be zero

\be
c_1 (\mathcal{L}_Y) \cdot c_1 (S) = 0
\ee 

\noindent If we further require the complete elimination of these exotics, then we must demand also that the sum has to be zero, so from Eq. (\ref{Ysum}), we can see that the line bundle has to satisfy

\be 
c_1 (\mathcal{L}_Y) \cdot c_1 (\mathcal{L}_Y) = -2 
\ee

\noindent which is equivalent to the condition for $c_1 (\mathcal{L}_Y)$ to correspond to a root of $E_8$.

\subsection{$E_6$ Bulk Exotics and their $SU(5)$ picture}

It has been shown in \cite{Beasley:2008kw} that bulk exotics coming from the adjoint of the GUT group on S cannot be avoided in the case where the gauge group is SO(10) or higher, and the breaking of the GUT group down to the Standard Model is achieved by flux breaking.  If we take the GUT group to be $E_6$, the spectrum can be found by decomposing under the $E_8$ enhancement

\begin{align}
\label{E6000}
\begin{split}
E_8 & \supset   E_6 \times SU(3)_{\perp} \\
248 & \rightarrow   (78,1)+(27,3)+(\overline{27},\overline{3})+(1,8)
\end{split} 
\end{align}

\noindent The SM can be achieved by turning on fluxes in the U(1)s contained in the following sequence of rank preserving breakings:

\begin{align}
 \label{E6001}
\begin{split}
E_6 & \rightarrow SO(10) \times U(1)_{\psi}  \\
& \rightarrow SU(5)\times  U(1)_{\chi} \times U(1)_{\psi}  \\
& \rightarrow SU(3) \times SU(2) \times U(1)_Y \times  U(1)_{\chi} \times U(1)_{\psi}
\end{split} 
\end{align}

%\subsection{The case of the adjoint of $E_6$}

\noindent In order to discuss the bulk exotics, we must decompose the adjoint of $E_6$ appearing in Eq. (\ref{E6000}) under the breaking pattern of Eq. (\ref{E6001}) as follows

\begin{align}
\label{78decomp}
\begin{split}
78 & \rightarrow (1,1)_{0,0,0} + \left\{(1,1)_{0,0,0} + (1,1)_{0,0,0} +(8,1)_{0,0,0} +(1,3)_{0,0,0} +(3,2)_{-5,0,0} + (\ov{3},2)_{5,0,0} \right.  \\ 
& \left. +(3,2)_{1,4,0} +(\ov{3},2)_{-1,-4,0}  +(\ov{3},1)_{-4,4,0} +(3,1)_{4,-4,0} + (1,1)_{6,4,0} +(1,1)_{-6,-4,0} \right\}  \\
& +\left\{(1,1)_{0,-5,-3}+(\ov{3},1)_{2,3,-3}+(1,2)_{-3,3,-3}+(1,1)_{6,-1,-3}+(3,2)_{1,-1,-3}+(\ov{3},1)_{-4,-1,-3} \right\} \\
& +\left\{(1,1)_{0,5,3}+(3,1)_{-2,-3,3}+(1,2)_{3,-3,3}+(1,1)_{-6,1,3}+(\ov{3},2)_{-1,1,3}+(3,1)_{4,1,3} \right\}
\end{split} 
\end{align}

\noindent All representations are charged under three U(1)s, and all triplets of U(1) charges can be expressed as a linear combination of the following line bundles

\begin{equation}
\mathcal{L}_1 = (5,0,0), \, \, \, \mathcal{L}_2 = (1,4,0), \, \, \, \mathcal{L}_3 = (1,-1,-3) \label{Bundles1}
\end{equation}

\noindent In Table~\ref{E6ex}  we write down the multiplicities of the exotic states coming from the adjoint of $E_6$ (where the correct normalisation for the $U(1)_Y$ is given by dividing by 6)
  \begin{table}[tbh] \centering%
\begin{tabular}{|ll||ll|}
\hline
 Exotic $X_i$& Multiplicity $n_i$& Exotic  $X_i$& Multiplicity  $n_i$\\
\hline $ X_1 = (\overline{3},2)_{\frac{5}{6}}$&$ n_1 = -\chi ({\cal L}_1, S) $ &
$X_6 = (\overline{3},1)_{\frac{1}{3}}$&$n_6 = -\chi ({\cal L}_2\otimes {\cal L}_3 , S)$\\
%\hline
$  X_2 = (3,2)_{\frac{1}{6}} $&$n_2 = -\chi ({\cal L}_2, S)$ &$X_7 = (1,2)_{-\frac{1}{2}} $&$n_{7} = -\chi ({\cal L}_1^{-1}\otimes {\cal L}_2\otimes {\cal L}_3 , S)$\\ 
%\hline 
$X_3 = (3,1)_{\frac{2}{3}} $&$n_3 = -\chi ({\cal L}_1^{-1}\otimes {\cal L}_2 , S)$&$X_8 = (1,1)_{1} $&$n_{8} = -\chi ({\cal L}_1\otimes {\cal L}_3 , S)$\\
%\hline 
$ X_4 = (1,1)_{1}$&$n_4 = -\chi ({\cal L}_1\otimes {\cal L}_2 , S)$&$ X_9 = (3,2)_{\frac{1}{6}}$ &$n_9=-\chi ( {\cal L}_3, S)$\\
%\hline 
$X_5 = (1,1)_{0} $&$ n_5 = -\chi ({\cal L}_2^{-1}\otimes {\cal L}_3 , S)$ &$X_{10} = (\ov{3},1)_{-\frac{2}{3}}$&$n_{10} = -\chi ({\cal L}_1^{-1}\otimes {\cal L}_3 , S)$\\
\hline
\end{tabular}%
\caption{$E_6$ bulk exotics and their multiplicities}
\label{E6ex}
\end{table}

\noindent We can see where the exotics fit into the SU(5) picture as follows (where the 
un-normalised $U(1)_Y\times  U(1)_{\chi} \times U(1)_{\psi}$ charges of the SU(5) states are 
indicated as subscripts),

\begin{alignat}{3}
\ov{5}_{3,-3} \rightarrow & (1,2)_{-3,3,-3} + && (\ov{3},1)_{2,3,-3} \nn \\
&X_7  && X_6 \label{5b}
\end{alignat} 
%\\
\begin{alignat}{4}
10_{4,0} \rightarrow & (1,1)_{6,4,0} + && (\ov{3},1)_{-4,4,0} + &&& (3,2)_{1,4,0} \nn \\
&X_4 && \ov{X}_3  &&& X_2 \label{101}
\end{alignat} 
%\\
\begin{alignat}{4}
10_{-1,-3} \rightarrow & (1,1)_{6,-1,-3} + && (\ov{3},1)_{-4,-1,-3} + &&& (3,2)_{1,-1,-3} \nn \\
&X_8  && X_{10}  &&& X_9 \label{102}
\end{alignat} 
%\\
\begin{alignat}{3}
24_{0,0} \rightarrow (1,1)_{0,0,0} +(8,1)_{0,0,0} +(1,3)_{0,0,0} + & (3,2)_{-5,0,0} + && (\ov{3},2)_{5,0,0} \nn \\
&\ov{X}_1  && X_1 \label{241}
\end{alignat} 

\subsection{Removing bulk exotics}

\noindent  When breaking the adjoint of a high gauge group there are always representations beyond those of the SM spectrum. These extraneous matter fields may be classified according to their charges in two categories: the ones that carry charges like the SM fields and those which have fractional  charges other than those of the SM quarks.  It can be seen that the exotics $\ov{X}_3$ and $X_{10}$ have the same SM quantum numbers as $u^c$, $X_2$ and $X_9$ have the same as $Q$, and $X_4$ and $X_8$ the same as $e^c$, with one set of states coming from Eq. (\ref{101}) and the other coming from Eq. (\ref{102}).  $X_1$ has exotic charges under the SM gauge group, and so we wish to remove these states from the spectrum.  $X_6$ and $X_7$ have the same SM quantum numbers as $d^c$ and $H_d$ respectively, and if present in the spectrum, we must require that they appear in vector pairs, and get mass via the couplings

\begin{align}
\label{5massterm}
\begin{split}
1_{0,0} \cdot \ov{5}_{3,-3} \cdot 5_{-3,3} & \rightarrow S X_6 \ov{X}_6 + S X_7 \ov{X}_7 \\
24_{0,0} \cdot \ov{5}_{3,-3} \cdot 5_{-3,3} & \rightarrow S \pr X_6 \ov{X}_6 + S \pr X_7 \ov{X}_7 
\end{split}
\end{align}

\noindent Requiring that $X_6$ and $X_7$ occur in vector pairs corresponds to imposing the conditions $n_6 - n_6 \cc = n_7 -n_7 \cc = 0$.  Using Table \ref{E6ex}, this leads to the following topological constraints

\begin{align}
c_1 (S) \cdot c_1 (\Lb) &= -c_1 (S) \cdot c_1 (\Lc) \label{const1} \\
c_1 (S) \cdot c_1 (\La) &= 0 \label{const2}
\end{align}

\noindent Presence of the $X_1$ states with exotic SM charges in the spectrum has been shown to lower the unification scale to unacceptable values, so requiring that these states are completely removed imposes the constraints (from Appendix \ref{A})

\begin{align}
\label{nox1}
\begin{split}
c_1 (S) \cdot c_1 (\La) &= 0  \\
c_1 (\La)^2 &= -2 
\end{split}
\end{align}

\noindent From Eq. (\ref{Bundles1}) and Table \ref{E6ex} along with the decompositions in the SU(5) picture, it can be seen that $\La$ corresponds to the hypercharge bundle.  As such, Eq. (\ref{nox1}) simply corresponds to the normal SU(5) condition $c_1 (\mathcal{L}_Y)^2 = -2$.  

If we were to impose that each type of exotic came in vector pairs individually (i.e $n_i = n_i \cc$ for i=1,...,10), from Appendix \ref{A} we would be led to the case of 
\be
c_1 (S) \cdot c_1 (\La) = c_1 (S) \cdot c_1 (\Lb) = c_1 (S) \cdot c_1 (\Lc) = 0
\ee
\noindent After imposing Eq. (\ref{nox1}), we can see that the only further choices we can make to eliminate some exotics (without getting negative numbers for any multiplicities) is

\begin{align}
c_1 (\Lb)^2 &= -2 \\
c_1 (\Lc)^2 &= -2
\end{align}

\noindent This ensures that the exotics $X_2$ and $X_9$ are completely removed, in addition to $X_1$.  All other exotics are present in vector pairs in this case, with their multiplicities given by  
\[    n_7 = n_7 \cc =2, \ \ \ \ n_i= n_i^* =1 \ \ \ \ ( i=3,4,5,6,8,10)  \]

%\begin{align}
%n_3 = n_3 \cc &= 1 \nn \\
%n_4 = n_4 \cc &= 1 \nn \\
%n_5 = n_5 \cc &= 1 \nn \\
%n_6 = n_6 \cc &= 1 \nn \\
%n_7 = n_7 \cc &= 2 \nn \\
%n_8 = n_8 \cc &= 1 \nn \\
%n_{10} = n_{10} \cc &= 1 \label{mult5}
%\end{align}

\subsection{A more general case}

As we have seen, we have two different 10 representations of SU(5), with different charges under $U(1)_{\psi}$ and $U(1)_{\chi}$, and so we can either give masses to the exotics contained in these 10s by couplings of the type

\begin{align}
1_{0,0} \cdot 10_{4,0} \cdot \ov{10}_{-4,0} & \rightarrow S X_2 \ov{X}_2 +S X_3 \ov{X}_3 + S X_4 \ov{X}_4 \nn \\
24_{0,0} \cdot 10_{4,0} \cdot \ov{10}_{-4,0} & \rightarrow S \pr X_2 \ov{X}_2 +S \pr X_3 \ov{X}_3 + S \pr X_4 \ov{X}_4 \label{massterm1}
\end{align}

\noindent or 

\begin{equation}
1_{-5,-3} \cdot 10_{4,0} \cdot \ov{10}_{1,3} \rightarrow X_5 (X_2 \ov{X}_9 ) +X_5 (\ov{X}_3 X_{10}) +X_5 (X_4 \ov{X}_8) \label{massterm2}
\end{equation}

\noindent where $X_5$ is a singlet exotic (corresponding to the `gluing morphism' of \cite{Donagi:2011dv}) coming from the 16 of SO(10), inside the 78 of $E_6$.  As such, we can more generally impose
\begin{align}
\begin{split}
n_2 + n_9 &= n_2 \cc + n_9 \cc \\
n_4 + n_8 &= n_4 \cc + n_8 \cc \\
n_3 \cc + n_{10} &= n_3 + n_{10} \cc
\end{split}
\end{align}

\noindent It can be seen that all three of these constraints are satisfied by imposing Eq. (\ref{const1}).  As such, with Eqs. (\ref{const2}, \ref{nox1}) also imposed, the multiplicities can be written in terms of the dot products

\begin{align}
A &= c_1 (S) \cdot c_1 (\Lb) = -c_1 (S) \cdot c_1 (\Lc) \nn \\
B &= c_1 (\Lb)^2 \nn \\
C &= c_1 (\Lc)^2 \label{dots1}
\end{align}

  \begin{table}[tbh] \centering%
\begin{tabular}{ll}
\hline
$n_1 = n_1^*= 0  $&$n_6 = -1 -\frac{B}{2} -\frac{C}{2} $ \\
$n_2 = -1 -\frac{A}{2} - \frac{B}{2}  $&$n_7 = -\frac{B}{2} -\frac{C}{2}$ \\
$n_3 = \frac{A}{2}-\frac{B}{2}       $&$n_8 = \frac{A}{2} -\frac{C}{2}$\\
$n_4 = -\frac{A}{2} - \frac{B}{2}          $&$n_9 = -1 +\frac{A}{2} -\frac{C}{2}$\\
$n_5 = -1 +A -\frac{B}{2} -\frac{C}{2}$&$n_{10} = \frac{A}{2} -\frac{C}{2}$ \\
\hline
\end{tabular}%
\caption{Multiplicities of the $E_6$ exotics in terms of the topological numbers $A,B,C$ (see text). }
\label{multi}
\end{table}

%\begin{alignat}{2}
%n_1 = n_1 \cc &= 0                              &&n_6 = -1 -\frac{B}{2} %-\frac{C}{2} \nn \\
%n_2 &= -1 -\frac{A}{2} - \frac{B}{2}            &&n_7 = -\frac{B}{2} %-\frac{C}{2} \nn \\
%n_3 &= \frac{A}{2}-\frac{B}{2}                  &&n_8 = \frac{A}{2} -\frac{C}{2} %\nn \\
%n_4 &= -\frac{A}{2} - \frac{B}{2}               &&n_9 = -1 +\frac{A}{2} %-\frac{C}{2} \nn \\
%n_5 &= -1 +A -\frac{B}{2} -\frac{C}{2} \, \, \, &&n_{10} = \frac{A}{2} %-\frac{C}{2} \label{mult6}
%\end{alignat}

\noindent The multiplicities are then given in Table~\ref{multi} 
where when dealing with the conjugate representations, A changes sign, but B and C keep the same sign.  We can now think about different combinations of exotic matter which satisfy these constraints, and consider the effect on gauge coupling unification.  The multiplicities of exotic matter are as follows 

\begin{alignat}{3}
n_Q &= n_2 + n_9 + n_2 \cc + n_9 \cc = -(B+C) -4 &&= \gamma -4 \nn \\
n_{u^c} &= n_3 + n_{10} + n_3 \cc + n_{10} \cc = -(B+C) &&= \gamma \nn \\
n_{e^c} &= n_4 + n_{8} + n_4 \cc + n_{8} \cc = -(B+C) &&= \gamma \nn \\
n_{d^c} &= n_6 + n_{6} \cc = -(B+C) -2 &&= \gamma -2 \nn \\
n_{H_d} &= n_7 + n_7 \cc = -(B+C) &&= \gamma  \label{mult7}
\end{alignat}

\noindent where we see that everything can be expressed in terms of the parameter $\gamma$, given in terms of Chern classes by

\begin{equation}
\gamma = -c_1 (\Lb)^2 - c_1 (\Lc)^2 \label{gamma}
\end{equation}

\noindent It can be seen from Table \ref{multi} that requiring $n_5 = n_5 \cc$ for the singlet $X_5$ leads us to the case $A=0$.  As such, all the exotic matter will satisfy $n_i = n_i \cc$, although we will still be able to get masses from both Eqs. (\ref{massterm1}) and (\ref{massterm2}).  It is important to note that as everything comes in conjugate pairs, anomalies are always cancelled.  We can now work out the contributions to the beta functions due to the exotic matter, and discuss gauge coupling unification.  Note that in order to satisfy the requirement that all multiplicities are positive, we must have $\gamma \geq 4$, with the minimal value being taken in the case where the line bundles satisfy the condition $c_1 (\Lb)^2 = c_1 (\Lc)^2 = -2$, meaning that $c_1 (\Lb)$ and $c_1 (\Lc)$ correspond to roots of $E_8$.  

\section{Gauge Coupling Unification}

In this section, the splitting of the gauge couplings at the GUT scale due to flux will be considered.  The eventual goal in the next section will be to study two types of semi-realistic model and include the effects of bulk exotics at various different mass scales.  Given the presence of these bulk exotics (plus the matter exotics in each model), the SM gauge couplings will be run up to the GUT scale, and it will be seen how closely they meet, and hence which models are consistent with the GUT scale relations.  In particular, it will be found that one MSSM like model will be consistent with GUT scale bulk exotics, and another E6SSM like model will be consistent with some bulk exotics at the TeV scale.

In this section, the effect of the bulk exotics alone on gauge coupling unification is discussed.  In $E_6$ models with flux breaking, we can view the breaking to the SM occurring in three steps.  Firstly, a flux along $U(1)_{\chi}$ breaks $E_6$ to $SO(10)$, then a flux along $U(1)_{\psi}$ breaks $SO(10)$ to $SU(5)$, and finally the hypercharge flux breaks $SU(5)$ to the SM.  Using Section 4 of \cite{Callaghan:2012rv}, it can be argued that in the $E_6$ and $SO(10)$ breakings, no relative splitting of the coupling constants occurs but only a constant shift, which at this level is unknown and can be treated as a free parameter of the model.  It is only the hypercharge flux which induces a relative splitting between the gauge couplings at the GUT scale, and so we will use the SU(5) relations at the GUT scale from \cite{Blumenhagen:2008aw}, effectively meaning that the aforementioned free parameter is set to zero.  It has been pointed out in \cite{Mayrhofer:2013ara} that the splitting relations of this type in F-theory are different from those in type IIB, and so we note that our treatment is simply an extrapolation of the type IIB results to F-theory.  In this way, we neglect corrections which can be important in the F-theory context, and so are working with an approximate treatment.

The remainder of this section just deals with bulk exotics for simplicity, while other exotics will be included in the analysis in the next section.  In essence, the gauge couplings are run from the MSSM values up to the GUT scale, and assuming bulk exotics at a mass scale $M_X$, constraints on the GUT scale and splitting of the couplings at this scale are studied.  This will provide the groundwork for the next chapter, where matter exotics will be considered in the spectrum as well (in the context of realistic models), and it will be seen which combinations of matter and bulk exotics are compatible with the constraints.

\subsection{The effect of bulk exotics at a single mass scale $M_X$}
It has been shown in \cite{Blumenhagen:2008aw} that in the context of an SU(5) GUT, the splitting at $M_{GUT}$ due to hypercharge flux is
\be\label{gcMU}
\begin{split}
\frac{1}{\alpha_3(M_{GUT})}&=\frac{1}{\alpha_G}-y
\\
\frac{1}{\alpha_2(M_{GUT})}&=\frac{1}{\alpha_G}-y+x
\\
\frac{1}{\alpha_1(M_{GUT})}&=\frac{1}{\alpha_G}-y+\frac 35 x
\end{split}
\ee

\noindent where $x=- \frac 12 {\rm Re} S\int c_1^2({\cal L}_Y)$ and  $y=\frac 12 {\rm Re} S\int c_1^2({\cal L}_a) $ associated with a non-trivial line bundle ${\cal L}_a$ and $S=e^{-\phi}+i\,C_0$ the axion-dilaton field.  It is argued in \cite{Callaghan:2012rv} that the $U(1)_{\psi}$ and $U(1)_{\chi}$ fluxes don't lead to any relative splittings of the gauge couplings at unification, although there could be a constant shift in all the couplings at each breaking.  This constant will be a free parameter of the model, and for simplicity we will set it to zero here.  As such, Eq. (\ref{gcMU}) can be used in the case of interest, and combining the three equations shows that the gauge couplings at $M_{GUT}$ are found to  satisfy  the  relation
  
%Combining these equations, the gauge couplings at $M_{GUT}$ are found to  satisfy  the  relation

\be
\frac{1}{\alpha_Y(M_{GUT})}=\frac 53 \,\frac{1}{\alpha_1(M_{GUT})}=\frac{1}{\alpha_2(M_{GUT})}+\frac 23 \frac{1}{\alpha_3(M_{GUT})}\label{SR2}
\ee

\noindent If we assume that the bulk exotics all decouple at a single mass scale $M_X$, the low energy values of the  gauge couplings are given by the evolution equations

\be\label{Brun}
\frac{1}{\alpha_a(M_Z)} = \frac{1}{\alpha_{a}(M_{GUT})}+\frac{b_a^x}{2\pi}\,\ln\frac{M_{GUT}}{M_X}+\frac{b_a}{2\pi}\,\ln\frac{M_{X}}{M_Z}
\ee

\noindent where $b_a^x$ are the beta functions above the scale $M_X$, and $b_a$ are the beta functions below this scale, i.e. those of the MSSM.  

Combining Eqs. (\ref{SR2}) and (\ref{Brun}) leads to the relation for the GUT scale

\be
M_{GUT} =  e^{\frac{2\pi}{\beta {\cal A}}\rho}\, \left(\frac{M_X}{M_Z} \right)^{1-\rho} M_Z
\label{M_U}
\ee
where ${\cal A}$ is a function of the experimentally known low energy values of the
SM gauge coupling constants
\ba
\frac{1}{\cal A} &=& \frac 53 \,\frac{1}{\alpha_1(M_Z)}-\frac{1}{\alpha_2(M_Z)}-\frac 23 \frac{1}{\alpha_3(M_Z)}
\nn\\
&=&\frac{\cos(2\theta_W)}{\alpha_{em}}-\frac 23 \frac{1}{\alpha_3(M_Z)}
\ea
Here use has been made of the relations $\alpha_Y=\alpha_e/(1-\sin^2\theta_W)$ and $\alpha_2=\alpha_e/\sin^2\theta_W$.
We have also introduced  the ratio $\rho$
\be
\rho  = \frac{\beta}{\beta_x}
\ee
where $\beta, \beta_{x}$ are the beta-function combinations in the regions $M_Z < \mu < M_{X}$  and $M_X < \mu < M_{GUT}$ respectively
\begin{align}
\beta_x &=b_Y^x-b_2^x-\frac 23b_3^x \label{betax2}\\
\beta &=b_Y-b_2-\frac 23b_3 \label{beta02}
\end{align}

Recall now the beta-function coefficients   ( $b_1=\frac 35\, b_Y$)
\ba
b_1&=& 6 +\frac{3}{10}(n_h+n_L)+\frac{1}{5}n_{d^c}+\frac{1}{10}n_Q+\frac{4}{5}n_{u^c} 
+\frac 35\,n_{e^c} \label{b1}
\\
b_2&=& \frac 12 (n_h+n_L)+ \frac 32\,n_Q \label{b2}
\\
b_3&=& -3+ \frac 12\,n_{d^c}+n_Q+\frac 12\,n_{u^c} \label{b3}
\ea
where $n_{h,L,...}$ counts the number of Higgses and exotic matter.

Below $M_{X}$ we have only the MSSM spectrum, thus $n_G=3,n_h=2$ and all extra matter contributions are zero, $n_i=0$, thus
\ba
\{b_Y,b_2,b_3\}=\{11,1,-3\}&\ra&\beta =b_Y-b_2-\frac 23b_3=12\nn
\ea

\noindent Ignoring possible matter curve exotics for now and just focusing on the bulk exotics, above $M_X$ we have the extra matter given in Eq. (\ref{mult7}) in addition to the two Higgses of the MSSM, giving for the beta functions 

\begin{align}
b_Y^x &= \frac{1}{3} (29+10 \gamma) \nn \\
b_2^x &= 2 \gamma - 5 \nn \\
b_3^x &= 2(\gamma - 4) \nn \\
\beta_x &= 20 \label{betax}
\end{align}

\noindent As such, we can see that the beta function combination $\beta_x$ doesn't depend on the parameter $\gamma$ and so the choice of this parameter will not affect the unification scale.  Putting the numbers into Eq. (\ref{M_U}) gives 

\be
M_{GUT} = \left( \frac{M_Z}{91\ \rm{GeV}}\right)^{\frac{3}{5}} 
 \left( \frac{M_X}{2.09 \times 10^{16}\ \rm{GeV}}\right)^{\frac{2}{5}}
 2.09 \times 10^{16} \ 
  \rm{GeV} \label{MGN}
\ee 

There are a number of points associated with the above equations that we would like to clarify.
Firstly we emphasise that Eq.~(\ref{gcMU}) applies not only for $SU(5)$ but also for $E_6$ models
for the reasons discussed above. Secondly we emphasise that the parameter $x$ in Eq.~(\ref{gcMU})
is constrained to be smaller than unity, since $x=- \frac 12 {\rm Re} S\int c_1^2({\cal L}_Y)$
and $\int c_1^2({\cal L}_Y)= -2$ (as discussed earlier) and ${\rm Re} S<1$.
This places a constraint on possible spectra consistent with unification, as discussed in the next section.
The above calculation of the GUT scale in Eq.~(\ref{M_U}) assumes such a compatible spectrum.

%Putting the numbers into Eq. (\ref{M_U}) gives 

%\be
%M_{GUT} = \left( \frac{M_Z}{91\ \rm{GeV}}\right)^{\frac{3}{5}} 
% \left( \frac{M_X}{2.09 \times 10^{16}\ \rm{GeV}}\right)^{\frac{2}{5}}
% 2.09 \times 10^{16} \ 
%  \rm{GeV} \label{MGN}
%\ee
  
%\noindent Clearly, if we take $M_X = 2.09 \times 10^{16} \rm{GeV}$, we also get $M_{GUT}=2.09 \times 10^{16} \rm{GeV}$.  We can see how different values of $M_X$ change the GUT scale in the graph of Figure \ref{MX}.

%\begin{figure}[!ht]
%\centering
%\includegraphics[scale=1,angle=0]{bulkexoticsgraph1.pdf}
%\caption{\small{Graph of how the bulk exotic mass scale $M_X$ impacts on the GUT scale $M_{GUT}$.}
%} \label{MX}
%\end{figure}

\subsection{The splitting parameter, x}

Combining Eqs. (\ref{gcMU}) and (\ref{Brun}) leads to the following expression for the parameter $x$

\begin{align}
x &= \left(\frac{1}{\alpha_2}-\frac{1}{\alpha_3}\right)_{M_{GUT}} \nn \\
&= \left(\frac{1}{\alpha_2}-\frac{1}{\alpha_3}\right)_{M_Z} + \frac{b_3^x-b_2^x}{2\pi}\log\left(\frac{M_{GUT}}{M_X}\right) + \frac{b_3-b_2}{2\pi}\log\left(\frac{M_{X}}{M_Z}\right) \nn \\
&= \frac{26 \sin^2 \theta_{W} - 3}{20 \alpha_{em}} - \frac{9}{10 \alpha_3 } - \frac{11}{10 \pi} \log\left(\frac{M_{X}}{M_Z}\right) \label{xsplit}
\end{align}

\begin{figure}[!ht]
\centering        
\includegraphics[scale=1,angle=0]{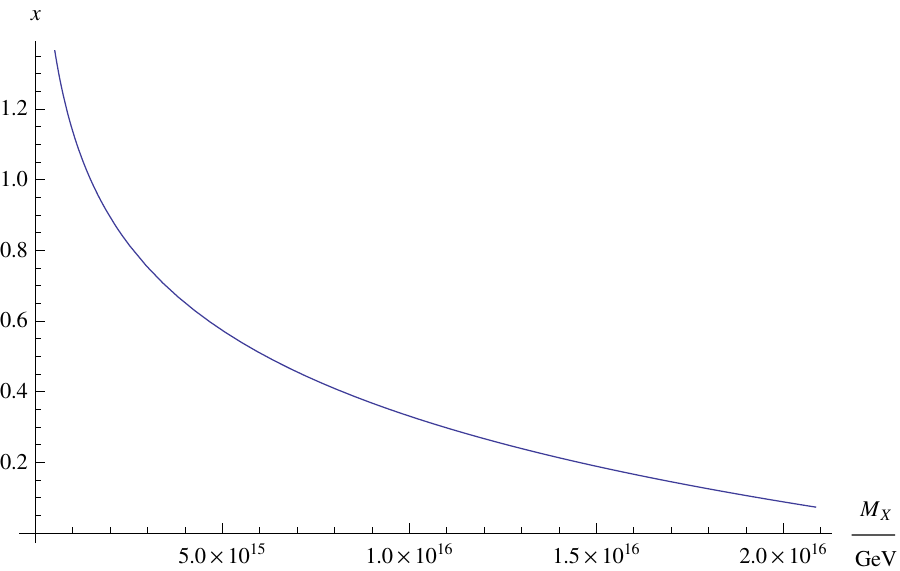}
\caption{The dependence of the splitting parameter x on the bulk exotic mass scale $M_X$.
Only values of $x\leq 1$ are acceptable, leading to the approximate lower bound on the bulk exotic
mass scale $M_X \geq 2 \times 10^{15} \ \rm{GeV}$.
Note that this bound assumes that no matter exotics are present.} \label{xg}
\end{figure}

\noindent The dependence of $x$  on the bulk exotic mass scale $M_X$ is shown in Figure \ref{xg}.  It can be seen that the splitting of the gauge couplings at the unification scale doesn't depend on the parameter $\gamma$.  It should also be noted that as x is given by $x=- \frac 12 {\rm Re} S\int c_1^2({\cal L}_Y)$ with $S=e^{-\phi}+i\,C_0$, it must take a value between 0 and 1.

\section{$E_6$ Models from F-theory}
\subsection{Matter exotics only}
We start by looking at the class of models proposed in \cite{Callaghan:2011jj}, which were motivated by the fact that any model involving complete 27s of $E_6$, with no matter coming from the adjoint 78 representation, automatically
satisfies anomaly cancellation involving most of the extra U(1)s.  \footnote{In reality, there is a small region of energy between $1.5 \times 10^{16} GeV$ and $M_{GUT} \approx 2 \times 10^{16} GeV$ where a particular anomaly is not cancelled, but the anomalies involving just the U(1)s inside $E_6$ are always cancelled.  There is a discussion of this subtle point in \cite{Callaghan:2011jj}.} Table \ref{1} shows the model building freedom we have in choosing the M and N integers specifying the flux breaking, and how these choices determine the Standard Model particle content of the model.  Here we make the same choices for the Ms and Ns as in \cite{Callaghan:2011jj} and these choices are summarised in Table \ref{1}. Although the SM particle content is equivalent to 
having three complete 27s, it is clear that the particles
are originating from incomplete multiplets of several different 27s.
The $U(1)_{N}$ charges of all the particles in the spectrum can be computed, and the results are shown in Table 3. As required, the right handed neutrinos have zero charge under $U(1)_{N}$. 
In Table \ref{1}, arbitrary numbers of singlets are allowed in the spectrum for now, so that we can calculate the restrictions on these numbers later on.  The final column of Table \ref{1} shows the low energy spectrum of the E6SSM obtained by eliminating the required exotics from the previous column, which shows the SM particle content after flux breaking.  By comparing the final two columns of Table \ref{1}, we can see that the matter exotics which we wish to remove are the vector pairs $2(L+\overline{L}), Q+\overline{Q}, 2(u^c+\overline{u^c}),d^{c}+\overline{d^{c}}$ and $H_d+\overline{H_d}$.  Large masses will be generated for these fields through their coupling to SM singlet fields which acquire large VEVs.     

From the $E_6$ point of view, the only $E_6$ allowed trilinear term in the superpotential is $27_{t_1} 27_{t_1} 27_{t_3}$.  The vectorlike pairs which we wish to remove from the low energy particle content are those which have components in both the $27_{t_{1}}$ and $27_{t_{3}}$ multiplets.  As such, they are removed by introducing $\theta_{31}$, an $E_6$ singlet, with couplings:
\beq
\theta_{31}27_{t_1'}\overline{27_{t_3'}}= \theta_{31}Q\overline{Q}+ \theta_{31}(2u^c)(2\overline{u^c})+
 \theta_{31}d^c\overline{d^c}+ \theta_{31}(2L)(2\overline{L})+
\theta_{31}H_d\overline{H_d}.
\label{31}
\eeq

If $\theta_{31}$ gets a large VEV  these vector states get large masses as required.  The difference between this case and model 1 \cite{Callaghan:2011jj} is that in model 1, $\theta_{34}$ also gets a large VEV.  This singlet has the following couplings
\beq
\theta_{34}5_1\overline{5_2}
=\theta_{34}[3D+2H_u][3\overline{D}+3H_d]=
\theta_{34}[3(D\overline{D})]+\theta_{34}[2(H_uH_d)].
\label{34}
\eeq
In the E6SSM, these matter exotics are light, and so instead of getting a large VEV, this singlet now must acquire a TeV scale VEV.  It was checked that the F and D-flatness constraints are satisfied, and that rapid proton decay is forbidden for the realisation of the spectrum \cite{Callaghan:2011jj}.

\begin{table}[ht]
\small
\begin{tabular}{|c|c|c|c|c|c|c|c|c|}
\hline
$E_6$ & $SO(10)$ & $SU(5)$  & Weight vector & $Q_N$ & $N_Y$ & $M_{U(1)}$ & SM particle content& Low energy spectrum\\
\hline
$27_{t_1'}$ & $16$ & $\overline{5}_3$ & $t_1+t_5$ & $\frac{1}{\sqrt{10}}$ & $1$ & $4$ &
$4d^c+5L$&$3d^{c}+3L$\\
\hline
$27_{t_1'}$ & $16$ & $10_M$ & $t_1$ & $\frac{1}{2\sqrt{10}}$ &$-1$ & $4$ &
$4Q+5u^c+3e^c$&$3Q+3u^{c}+3e^{c}$\\
\hline
$27_{t_1'}$ & $16$ & $\theta_{15}$ & $t_1-t_5$ & 0 & $0$ & $n_{15}$ &
$3\nu^c$&-\\
\hline
$27_{t_1'}$ & $10$ & $5_1$ & $-t_1-t_3$ & $-\frac{1}{\sqrt{10}}$ &$-1$ & $3$ &
$3D+2H_u$&$3D+2H_u$\\
\hline
$27_{t_1'}$ & $10$ & $\overline{5}_2$ & $t_1+t_4$ & $-\frac{3}{2\sqrt{10}}$ &$1$ & $3$ &
$3\overline{D}+4H_d$&$3\overline{D}+3H_d$\\
\hline
$27_{t_1'}$ & $1$ & $\theta_{14}$ & $t_1-t_4$ & $\frac{5}{2\sqrt{10}}$ &$0$ & $n_{14}$ &
$\theta_{14}$& $\theta_{14}$\\
\hline
$27_{t_3'}$ & $16$ & $\overline{5}_5$ & $t_3+t_5$ & $\frac{1}{\sqrt{10}}$ & $-1$ & $-1$ &
$\overline{d^c}+2\overline{L}$&-\\
\hline
$27_{t_3'}$ & $16$ & $10_2$ & $t_3$ & $\frac{1}{2\sqrt{10}}$ &$1$ & $-1$ &
$\overline{Q}+2\bar{u^c}$&-\\
\hline
$27_{t_3'}$ & $16$ & $\theta_{35}$ & $t_3-t_5$ & 0& $0$ & $n_{35}$ &
$-$&-\\
\hline
$27_{t_3'}$ & $10$ & $5_{H_u}$ & $-2t_1$ & $-\frac{1}{2\sqrt{10}}$ &$1$ & $0$ &
$H_u$&$H_{u}$\\
\hline
$27_{t_3'}$ & $10$ & $\overline{5}_4$ & $t_3+t_4$ & $-\frac{3}{2\sqrt{10}}$ &$-1$ & $0$ &
$\overline{H_d}$&-\\
\hline
$27_{t_3'}$ & $1$ & $\theta_{34}$ & $t_3-t_4$ & $\frac{5}{2\sqrt{10}}$ & $0$ & $n_{34}$ &
2$\theta_{34}$& 2$\theta_{34}$ \\
\hline
- & $1$ & $\theta_{31}$ & $t_3-t_1$ & 0 & $0$ & $n_{31}$ &
$\theta_{31}$&-\\
\hline
- & $1$ & $\theta_{53}$ & $t_5-t_3$ & 0 & $0$ & $n_{53}$ &
$\theta_{53}$&-\\
\hline
- & $1$ & $\theta_{54}$ & $t_5-t_4$ & $\frac{5}{2\sqrt{10}}$ &$0$ & $n_{54}$ &
$\theta_{54}$&-\\
\hline
- & $1$ & $\theta_{45}$ & $t_4-t_5$ & $-\frac{5}{2\sqrt{10}}$ &$0$ & $n_{45}$ &
$\theta_{45}$&-\\
\hline
\end{tabular}
\caption{\small Complete $27$s of $E_6$ and their $SO(10)$ and $SU(5)$ decompositions.
The $SU(5)$ matter states decompose into SM states as
$\overline{5}\rightarrow d^c,L$ and $10\rightarrow Q,u^c,e^c$ with right-handed neutrinos
$1\rightarrow \nu^c$, while $SU(5)$ Higgs states decompose as $5\rightarrow D,H_u$ and
$\overline{5}\rightarrow \overline{D},H_d$, where $D, \overline{D}$ are exotic colour triplets and antitriplets.
We identify RH neutrinos as $\nu^c=\theta_{15}$.  Arbitrary singlets are included for giving mass to neutrinos and exotics and to ensure F- and D- flatness.}
\label{1}
\end{table}%

Clearly the matter exotics ($d+ \ov{d}^c$), ($Q+ \ov{Q}$), ($H_d + \ov{H}_d$), 2($L+ \ov{L}$), 2($u^c + \ov{u}^c$) get masses and decouple at some scale $M_{\theta_{31}} < M_{GUT}$ due to the couplings in Eq. (\ref{31}).  The matter exotics 3($D+ \ov{D}$), 2($H_u , H_d$) get masses and decouple at a scale $M_{\theta_{34}} < M_{\theta_{31}}$ due to the couplings in
Eq. (\ref{34}). 
% The GUT scale is determined to be $M_{GUT} = 2.09 \times 10^{16} \rm{GeV}$ in both models. 
%$M_{\theta_{31}}$ is computed by requiring D-flatness of the models, and in both cases it is close to the GUT scale.
In \cite{Callaghan:2011jj, Callaghan:2012rv} (which we will call models 1 and 2 respectively from now on) two different classes of model were proposed only distinguished by the mass scales of the matter exotics.
The scales of the two models are summarised below.

In model 1 (``MSSM''):
\begin{align}
M_{\theta_{31}}^{(1)} &= 1.31 \times 10^{16} \rm{GeV} \nn \\
M_{\theta_{34}}^{(1)} &= 0.306 \times 10^{16} \rm{GeV} \nn
\end{align}

In model 2 (``E6SSM''):
\begin{align}
M_{\theta_{31}}^{(2)} &= 1.44 \times 10^{16} \rm{GeV} \nn  \\
M_{\theta_{34}}^{(2)} &= 1 \times 10^{3} \rm{GeV} \nn 
\end{align}
The main difference between the two models is clearly that in model 1 the $\theta_{34}$ 
matter exotics are computed to be almost as heavy as the $\theta_{31}$ exotics, whereas in model 2 the $\theta_{34}$ matter exotics are kept light, getting TeV scale masses. 
We see that model 1 reproduces the MSSM somewhat below the GUT scale
since only the MSSM spectrum survives below 
$M_{\theta_{34}}$, whereas model 2 corresponds to the so called E6SSM above the TeV scale
(or NMSSM+ if the $U(1)_{N}$ gauge group is broken at high energy).
However strictly speaking the spectrum of model 2 is not quite that of the E6SSM since it only contains the matter
content of three 27 dimensional representations of $E_6$ and does not contain the extra vector-like
matter usually denoted as $H'$ and $\overline{H'}$ which is required for gauge coupling unification.
As we shall see shortly, the role of the extra $H'$ and $\overline{H'}$ will be played by bulk exotics.

\subsection{The complete spectra of potential models}

In the previous section, we have reviewed the spectra of the models presented in \cite{Callaghan:2011jj} and \cite{Callaghan:2012rv}.  In order to include bulk exotics into these cases, we can note from Eq. (\ref{mult7}) that taking the minimal case of $\gamma = 4$ leads to the following vector pairs of bulk exotics, which have to be added to the spectrum of any chosen model:

\be
2(u^c + \ov{u^c}) + 2(e^c + \ov{e^c}) + 2(H_d + \ov{H_d}) + (d^c + \ov{d^c}) \label{vecpairsmin}
\ee

\noindent The question now remains as to what masses these vector-like exotics acquire.  Those exotics which originate from 5 representations at the SU(5) level become massive through the couplings in Eq. (\ref{5massterm}).  The same singlets which appear in this equation and acquire VEVs can also give mass to the bulk exotics from 10s of SU(5), through Eq. (\ref{massterm1}).  However, the 10-like exotics can also get masses through the coupling in Eq. (\ref{massterm2}).  This gives three distinct possibilities for the masses of the bulk exotics: 
(i) all can get masses at the GUT scale, 
(ii) all can get TeV scale masses, or
(iii) the 5-like exotics could acquire TeV scale masses, while the 10-like ones could acquire GUT scale masses through the coupling of Eq. (\ref{massterm2}). 
\noindent All of these possibilities will be discussed in the context of different models in this section.  

It should be noted that in terms of anomaly cancellation, we will end up with two cases: one where we have the MSSM at the TeV scale and all exotics near the GUT scale, and one where we have the E6SSM and some extra vector pairs of bulk exotics at the TeV scale, and everything else at the GUT scale.  Clearly the MSSM case is anomaly free at the TeV scale and so we simply must check that anomalies are cancelled by the spectrum as a whole.  Not including the bulk exotics for now, Appendix \ref{C} shows that all anomalies are cancelled apart from a $U(1)_Y \times U(1)_{\perp}^2$ anomaly.  However, this anomaly only exists in the small region of energy between the $\theta_{31}$ VEV and the GUT scale, and due to the errors in our energy scale estimates, this shouldn't be a problem. 
In the E6SSM case, clearly the whole spectrum suffers the same issue as the first case, and so we must just check the TeV spectrum.  Clearly, if we just had the low energy matter of the E6SSM we would be free from anomalies due to the presence of complete 27s of E6.  As such, it is only the bulk exotics at the TeV scale which could possibly spoil this fact.  However, these bulk exotics come in vector pairs and so they come with a built in minus sign between the $D_{abc}$ anomaly coefficients, thereby cancelling all anomalies at the TeV scale.  It should also be noted that \cite{Mayrhofer:2013ara} points out that anomaly cancellation constraints can be relaxed in the case of geometrically massive U(1)s in F-theory.

\subsection{High scale bulk exotics }

The above analysis does not so far include the effect of bulk exotics. However, as we have seen
earlier in the paper, such bulk exotics are an inevitable consequence of the flux breaking of $E_6$.
As remarked above, such additional bulk exotics at the TeV scale, not included in the spectrum
so far, are able to provide the extra vector-like matter to enable gauge unification to be achieved for the E6SSM.
However the resulting spectrum will differ somewhat from that of the E6SSM, providing a distinctive experimental
signature and providing a smoking gun test of the F-theory model at the LHC.
\begin{figure}[!ht]
\centering       
\includegraphics[scale=1,angle=0]{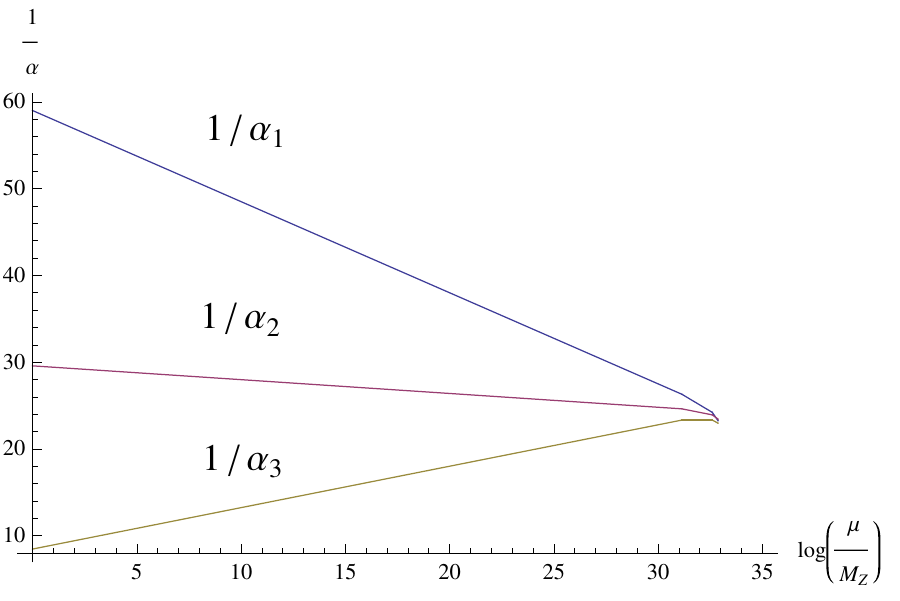}       
\caption{Gauge coupling unification in model 1 (MSSM) with high scale bulk exotics.}
\label{bgraphs}
\end{figure}

\begin{figure}[!ht]
\centering       
\includegraphics[scale=1,angle=0]{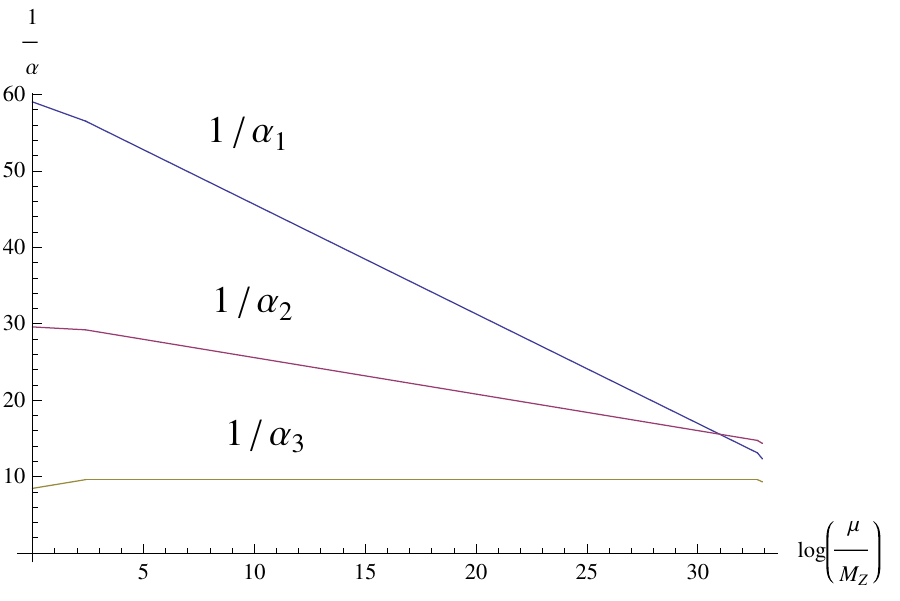}       
\caption{Gauge couplings fail to unify in model 2 (E6SSM) with high scale bulk exotics.}
\label{m1bgraph}
\end{figure}

In both the above models, the beta function combination given in Eq. (\ref{beta02}) is given by $\beta = 12$ (the MSSM value) in all of the regions $M_Z < \mu < M_{\theta_{34}}$ and $M_{\theta_{34}} < \mu < M_{\theta_{31}}$ and $M_{\theta_{31}} < \mu < M_{GUT}$.  As such, assuming that the bulk exotics get masses $M_X$, somewhere between $M_{\theta_{31}}$ and $M_{GUT}$, we will have an equation analogous to Eq. (\ref{M_U})

\be
M_{GUT} =  e^{\frac{2\pi}{\beta {\cal A}}\rho}\, M_Z^{\rho} M_{\theta_{34}}^{\eta - \rho} M_{\theta_{31}}^{\lambda-\eta} M_{X}^{1-\lambda}
\label{M_U2}
\ee

\noindent where in the same way as for Eq. (\ref{M_U}), $\rho = \eta = \lambda = \frac{3}{5}$.  As such, the GUT scale only depends on the mass of the bulk exotics, and is still given by Eq. (\ref{MGN}).  If we take $M_X =M_{GUT}$, the RGE analysis is obviously unchanged from that of \cite{Callaghan:2011jj, Callaghan:2012rv}, however if we take $M_X = M_{\theta_{31}}$, the GUT scale is lowered slightly by Eq. (\ref{MGN})

\begin{align}
M_{GUT}^{(1)} &= 1.73 \times 10^{16} \rm{GeV} \nn \\
M_{GUT}^{(2)} &= 1.80 \times 10^{16} \rm{GeV} \nn 
\end{align}

For model 1 (MSSM) the one loop running of the couplings is shown in Figure \ref{bgraphs}.  
This takes into account the modification of the beta functions due to the bulk exotics above the scale $M_X = M_{\theta_{31}}$.
In this case the couplings are split by $2 \%$ (compared to $1.3 \%$ when the bulk exotics are not taken into account), and it can be seen that the effect of bulk exotics near the GUT scale on the splitting of the gauge couplings is small ($0.5-1 \%$ depending on the model). 

For model 2  (E6SSM) the splitting is $35 \%$ (compared to $34.5 \%$ in the case with no bulk exotics), which would correspond to $x \sim 5$. This is shown in Figure \ref{m1bgraph}. 
As pointed out before, x must take a value between 0 and 1 and so model 2 must be ruled out in the case where the bulk exotics get masses near the GUT scale.

\subsection{Low scale bulk exotics}

We have seen that as long as the bulk exotics get masses close to the GUT scale, the GUT scale is not lowered drastically.  However, due to the fact that the bulk exotic spectrum ensures anomaly cancellation, the gauge groups $U(1)_{\chi}$ and $U(1)_{\psi}$ and the bulk exotics could in principle survive to the TeV scale.  We will now look at this possibility that at least some of the bulk exotics are light.  From Eqs. (\ref{b1}, \ref{b2}, \ref{b3}) we have

\begin{align}
\beta &= 12 +n_{u^c} + n_{e^c} - 2 n_Q \nn \\
\delta \beta &= \delta n_{u^c} + \delta n_{e^c} - 2 \delta n_Q \label{deltab}
\end{align}

\noindent where $\delta \beta = \beta_x - \beta$ is the difference in $\beta$ as we move a higher energy scale where a number of exotics ($\delta n_{u^c}$, $\delta n_{e^c}$ and $\delta n_Q$) join with the massless spectrum.  In models 1 and 2 there is no exotic $e^c$ type matter and the only $Q$ and $u^c$ exotics get the same mass, near the GUT scale.  In both models, there is twice as much $u^c$-like exotic matter as there is Q-like, and so $\delta \beta = 0$ when we don't take into account contributions from the bulk exotics.  For the bulk exotics, Eq. (\ref{mult7}) gives 

\be
\delta \beta = n_{u^c} + n_{e^c} - 2 n_Q = 8 
\ee

\noindent Previously, we looked at the case where $M_X \geq M \pr$ and we found that the GUT scale is slightly lowered.  If we now consider the case where $M_X \leq M_{\theta_{34}}$, Eq. (\ref{M_U2}) gets modified to 

\begin{equation}
M_{GUT} =  e^{\frac{2\pi}{\beta {\cal A}}\rho}\, M_Z^{\rho} M_{X}^{\eta - \rho} M_{\theta_{34}}^{\lambda-\eta} 
M_{\theta_{31}}^{1-\lambda} \label{lowbulk}
\end{equation}

\noindent with 

\begin{align}
\rho &= \frac{\beta}{\beta_{\theta_{31}}} = \frac{3}{5} \nn \\
\eta &= \frac{\beta_{x}}{\beta_{\theta_{31}}} = 1 \nn \\
\lambda &= \frac{\beta_{\theta_{34}}}{\beta_{\theta_{31}}} = 1 
\end{align}

\begin{figure}[!ht]
\centering       
\includegraphics[scale=1,angle=0]{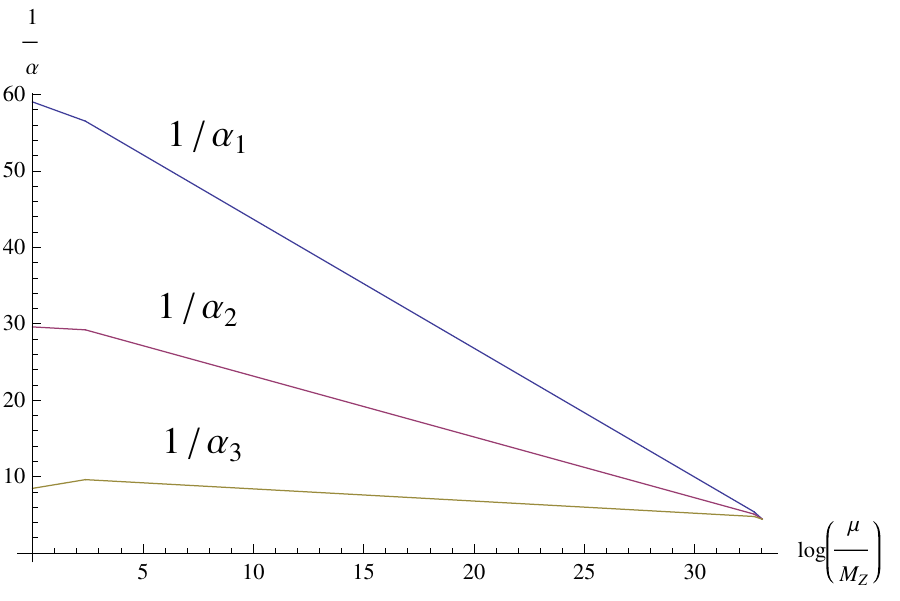}       
\caption{Gauge coupling unification in model 2 (E6SSM) with TeV scale bulk exotics.}
\label{model2uni}
\end{figure}

\noindent Again, we end up with Eq. (\ref{MGN}) for the GUT scale, with the bulk exotic mass (the mass of those coming from a 10 of SU(5) if we allow the 5s and 10s to get different masses) being the only exotic mass entering the equation.  As such, apart from the possibility that all bulk exotics get masses near the GUT scale (as previously discussed), we have two other possibilities:

\begin{itemize}
\item \emph{All bulk exotics at the TeV scale:} In this case Eq. (\ref{lowbulk}) tells us that $M_{GUT} \sim 1 \times 10^{11} \rm{GeV}$.  It may seem at first sight that such a low unification scale would lead to dangerous dimension 6 operators giving proton decay rates which are much faster than experimentally observed.  However, in \cite{Ibanez:2012zg} a method has been pointed out for suppressing proton-decay in F-theory SU(5) with hypercharge flux breaking.  The idea is that since the dangerous operators involve the SU(5) gauge bosons X,Y in trilinear couplings such as $X Q u^c$, a computation would consist of firstly computing the trilinear coupling by using the wavefunction overlap techniques of eg. \cite{Camara:2011nj}, and then integrating out X,Y.  The key is that the SU(5) gauge bosons need not be localised on a matter curve, but can be spread out over S.  As such, these fields feel the effect of hypercharge flux in a different way to those on matter curves, and this gives rise to a suppression of the integral.  This way, we can in principle avoid rapid proton decay, even with a seemingly low unification scale.  Even though this is the case, when all the bulk exotics are at the TeV scale the splitting of the gauge couplings is large, and $x > 1 $.  As such, this possibility must be ruled out and we must look at the next case.

\item \emph{Bulk exotics from 10s heavy, but those from 5s light:} As the singlets $S$ and $S \pr$ which give the 5 state mass through Eq. (\ref{5massterm}) can also give the 10s mass through Eq. (\ref{massterm1}), we reject the possibility of heavy 5s and light 10s.  However, since the 10s can get mass from a different singlet in Eq. (\ref{massterm2}), it would seem that there is a possibility of giving this singlet a much bigger VEV, and keeping the 10s heavy whilst the 5s could be TeV exotics.  If this was the case, we would once again have $M_{GUT} \sim 2 \times 10^{16} \rm{GeV}$ due to the fact that the 5s don't contribute to $\delta \beta$ of Eq. (\ref{deltab}).  If the splitting parameter x is calculated for this case with the spectrum of model 1, it turns out to be negative so again we must rule this case out. This means that for model 1, high energy bulk exotics are the only possibility, but on the contrary we will see that for model 2 these low energy bulk exotics are the only possibility.
As pointed out previously, model 2 which has TeV scale exotics in it's spectrum cannot be compatible with bulk exotics with masses close to the GUT scale, as $x > 1$ which is forbidden.  However, if we have the bulk exotics which belong to 5s of SU(5) at the TeV scale as described above, it turns out that the multiplicities of exotic states forced upon us by topological constraints make the couplings unify.  If we take the mass of the exotics from 10s to be $M_{GUT}$, we find $x \sim 0.01$, corresponding to a splitting of approximately $0.2 \%$.  This effect is illustrated in Figure \ref{model2uni}, which shows how the low energy bulk exotics are precisely what is needed to make the couplings unify.  In addition to the 3($D+ \ov{D}$), 2($H_u , H_d$) exotics which are also at the TeV scale, this leads to a characteristic spectrum involving TeV vector-like pairs of $d^c$ and $H_d$ exotics, with the distinguishing feature that there will always be one more vector pair of $H_d$ states than $d^c$s.  (In the $\gamma=4$ case, we have one pair of $d^c$ states and two pairs of $H_d$ states).
The low energy spectrum of this model is summarised in Table \ref{tevbulkspectrum}
\end{itemize}

\begin{table}[ht]
\begin{center}
\small
\begin{tabular}{|c|c|c|c|}
\hline
$E_6$ origin & $SU(5)$ origin & TeV scale spectrum & $U(1)_N$ \\ 
\hline
$27_{t_1'}$ & $\ov{5}$ & $3d^{c}+3L$ & $\frac{1}{\sqrt{10}}$ \\
\hline
$27_{t_1'}$ & $10$ & $3Q+3u^{c}+3e^{c}$ & $\frac{1}{2 \sqrt{10}}$ \\
\hline
$27_{t_1'}$ & $5$ & $3D+2H_u$ & $-\frac{1}{\sqrt{10}}$ \\
\hline
$27_{t_1'}$ & $\ov{5}$ & $3\overline{D}+3H_d$ & $-\frac{3}{2 \sqrt{10}}$ \\
\hline
$27_{t_1'}$ & $1$ & $\theta_{14}$ & $\frac{5}{2 \sqrt{10}}$ \\
\hline
$27_{t_3'}$ & $5$ & $H_u$ & $-\frac{1}{2 \sqrt{10}}$ \\
\hline
$27_{t_3'}$ & $1$ & $2 \theta_{34}$ & $\frac{5}{2 \sqrt{10}}$ \\
\hline
$78$ & $\ov{5}$ & $2X_{H_d} + X_{d^c}$  & $-\frac{3}{2 \sqrt{10}}$ \\
\hline
$78$ & $5$ & $2\ov{X}_{\ov{H}_d} +  \ov{X}_{\ov{d^c}}$  & $\frac{3}{2 \sqrt{10}}$ \\
\hline
\end{tabular}
\caption{\small The complete low energy spectrum for the E6SSM-like model with TeV scale bulk exotics.
The fields $Q$, $u^c$, $d^c$, $L$, $e^c$ represent quark and lepton SM superfields in the usual notation.
In this spectrum there are three families of $H_u$ and $H_d$ Higgs superfields, as compared to a single one in the MSSM. There are also three families of exotic $D$ and $\overline{D}$ colour triplet superfields,
where $\overline{D}$ has the same SM quantum numbers as $d^c$, and $D$ has opposite quantum numbers.
We have written the bulk exotics as $X$ with a subscript that indicates the SM quantum numbers of that state.
The superfields $\theta$ are SM singlets. }
\label{tevbulkspectrum}
\end{center}
\end{table}%

In the presence of large VEVs for $X_5$ and $\ov{X}_5$, the F and D flatness equations of \cite{Callaghan:2012rv} must be modified accordingly.  It is shown in Appendix \ref{B} that there is a solution to the flatness relations for this model where $X_5$ and $\ov{X}_5$ get large VEVs without giving rise to dangerous operators.  In this section we have taken $\left\langle X \right\rangle = M_{GUT}$ for simplicity and to illustrate it's effects.   

\section{Conclusions}

We have considered gauge coupling unification in $E_6$ F-Theory Grand Unified Theories
(GUTs) where $E_6$ is broken to the 
Standard Model (SM) gauge group using fluxes. In such models there are two types of exotics that can affect gauge coupling unification, namely matter exotics from the matter curves in the 27 dimensional representation of $E_6$ and the bulk exotics from the 
adjoint 78 dimensional representation of $E_6$.
The matter exotics have been considered previously in \cite{Callaghan:2011jj, Callaghan:2012rv}
where two models were considered which we called model 1 (MSSM) and model 2 (E6SSM),
which only differ by the mass scale of the matter exotics. In particular model 2 (E6SSM) involves
TeV scale matter exotics, equivalent to having complete 27 dimensional representations at the TeV scale,
and is inconsistent with gauge coupling unification.

We have mainly focussed on the question of bulk exotics arising from the flux breaking
of $E_6$, which have not previously been considered in the literature. 
We have explored the conditions required for either the complete or partial
removal of bulk exotics from the low energy spectrum.
In particular we have examined the conditions for the removal from the low energy spectrum of bulk exotic matter from the adjoint of $E_6$ in terms of topological properties of the manifold.  These conditions led to the fact that all vector-like pairs come in multiplicities which depend on one topological parameter, $\gamma$.  We studied how the bulk exotics affect the one loop RGE anaylsis, and it was shown that both the GUT scale and the splitting of the gauge couplings depend on the mass of the exotics, but not on $\gamma$, meaning that the results are general for any $E_6$ F-theory model using fluxes to break the GUT group.  

We then considered the effect of the necessary bulk exotics on gauge coupling unification.
For the case of high scale bulk exotics at a single mass scale $M_X$, 
where we assumed them to be the only type of exotics present,
we found that the requirement that they do not lead to an unacceptable splitting of gauge couplings
led to the bound $M_X \geq 2 \times 10^{15} \ \rm{GeV}$. We also found that having bulk exotics also led
to a lowering of the GUT scale down to about $M_X \geq 8 \times 10^{15} \ \rm{GeV}$ for the case where
the bound is saturated, $M_X = 2 \times 10^{15} \ \rm{GeV}$.

Finally we considered the effect of bulk exotics on the realistic $E_6$ models where also matter exotics
are present. For model 1 (MSSM) with high scale bulk exotics we found that unification is maintained
with $M_{GUT}$ being lowered slightly ($<15 \%$), and the splitting of the gauge couplings increased by less than $1\%$. Model 2 (E6SSM), which involves TeV scale matter exotics, is not much affected by the high scale bulk exotics and it still fails to provide gauge coupling unification. However for the case of low scale bulk exotics
the picture changes dramatically.

We considered the possibility that the bulk exotics from 5s of SU(5) could get TeV scale masses whereas those from 10s could be near the GUT scale due to a large VEV for a singlet charged under $U(1)_{\psi}$ and $U(1)_{\chi}$,
and showed that this possibility is consistent with the F and D flatness conditions.
The TeV scale bulk exotics then consist of vector-like pairs of $d^c$-like and $H_d$-like exotics, with the distinguishing feature that there will always be one more vector pair of $H_d$ states than $d^c$s.
In the case of model 2 (E6SSM), we found the elegant result that such bulk exotics 
combine with the TeV scale matter exotics to ensure almost perfect gauge coupling unification.

In summary, in $E_6$ models broken by flux, while it is possible that all bulk exotics as well
as matter exotics could have masses close to the GUT scale leading to 
acceptable gauge coupling unification with an MSSM type theory 
somewhat below the GUT scale, it is equally likely to have TeV scale exotics in such models.
We have discussed a remarkable possibility, namely model 2 (E6SSM) where the matter exotics
correspond to having complete 27 dimensional representations of $E_6$ at the TeV scale,
in which gauge coupling unification would fail badly without the presence of bulk exotics.
However including bulk exotics from the 5s of SU(5) at the TeV scale,
with those from the 10s near the GUT scale, restores gauge unification for this model.
We find this result remarkable, indeed almost miraculous, since the origin of the matter and bulk exotics
is apparently quite different in F-theory. 

We emphasise that, without such bulk exotics,
the TeV scale matter exotics of model 2 (E6SSM) would lead to an unacceptable splitting of the couplings,
and it is only the combination of TeV scale matter exotics from the 27s plus TeV scale 
bulk exotics from the 78 which restores gauge coupling unification.
The resulting TeV scale matter exotics plus bulk exotics is equivalent 
to having four extra $5+\overline{5}$ vector pairs of SU(5),
beyond the minimal supersymmetric standard model (MSSM) spectrum. 
This may be compared to the equivalent of three extra $5+\overline{5}$ vector pairs predicted by the E6SSM \cite{King:2005jy,King:2005my}.
Clearly the discovery of such exotics at the LHC would provide evidence for $E_6$ F-theory GUTs broken by flux.

\section*{Acknowledgements}

We are grateful to Graham Ross for discussions. 
We acknowledge the EU
ITN grant UNILHC 237920.
SFK acknowledges the EU
ITN grant INVISIBLES 289442
as well as support from the STFC Consolidated ST/J000396/1 grant.

\newpage

\section*{Appendix}

\begin{appendix}

\section{\label{A} Topological relations arising from the elimination of bulk exotics}

The requirement that each type of exotic matter occurs in vector pairs is given by $n_j - n_j \cc =0$.  The extra requirement which would mean that this type of exotic is completely eliminated from the spectrum is $n_j + n_j \cc =0$.  There requirements are given here for each type of exotic.  Note that all of these relations can't be satisfied at once, and are written here on the assumption that a subset of them will be satisfied.  

\begin{align}
n_1 - n_1 \cc = 0 & \Rightarrow c_1 (S) \cdot c_1 (\La) = 0 \\
n_1 + n_1 \cc = 0 & \Rightarrow c_1 (\La)^2 = -2
%\end{align}
\\
%\begin{align}
n_2 - n_2 \cc = 0 & \Rightarrow c_1 (S) \cdot c_1 (\Lb) = 0 \\
n_2 + n_2 \cc = 0 & \Rightarrow c_1 (\Lb)^2 = -2
%\end{align}
\\
%\begin{align}
n_3 - n_3 \cc = 0 & \Rightarrow c_1 (S) \cdot c_1 (\La) = c_1 (S) \cdot c_1 (\Lb) \\
n_3 + n_3 \cc = 0 & \Rightarrow c_1 (\La)^2 + c_1 (\Lb)^2= -2
%\end{align}
\\
%\begin{align}
n_4 - n_4 \cc = 0 & \Rightarrow c_1 (S) \cdot c_1 (\La) = - c_1 (S) \cdot c_1 (\Lb) \\
n_4 + n_4 \cc = 0 & \Rightarrow c_1 (\La)^2 + c_1 (\Lb)^2= -2
%\end{align}
\\
%\begin{align}
n_5 - n_5 \cc = 0 & \Rightarrow c_1 (S) \cdot c_1 (\Lc) = c_1 (S) \cdot c_1 (\Lb) \\
n_5 + n_5 \cc = 0 & \Rightarrow c_1 (\Lb)^2 + c_1 (\Lc)^2= -2
%\end{align}
\\
%\begin{align}
n_6 - n_6 \cc = 0 & \Rightarrow c_1 (S) \cdot c_1 (\Lb) = - c_1 (S) \cdot c_1 (\Lc) \\
n_6 + n_6 \cc = 0 & \Rightarrow c_1 (\Lb)^2 + c_1 (\Lc)^2= -2
%\end{align}
\\
%\begin{align}
n_7 - n_7 \cc = 0 & \Rightarrow c_1 (S) \cdot c_1 (\La) =  c_1 (S) \cdot c_1 (\Lb) + c_1 (S) \cdot c_1 (\Lc) \\
n_7 + n_7 \cc = 0 & \Rightarrow c_1 (\La)^2 + c_1 (\Lb)^2 + c_1 (\Lc)^2 = -2
%\end{align}
\\
%\begin{align}
n_8 - n_8 \cc = 0 & \Rightarrow c_1 (S) \cdot c_1 (\La) = - c_1 (S) \cdot c_1 (\Lc) \\
n_8 + n_8 \cc = 0 & \Rightarrow c_1 (\La)^2 + c_1 (\Lc)^2= -2
%\end{align}
\\
%\begin{align}
n_9 - n_9 \cc = 0 & \Rightarrow c_1 (S) \cdot c_1 (\Lc) = 0 \\
n_9 + n_9 \cc = 0 & \Rightarrow c_1 (\Lc)^2 = -2
%\end{align}
\\
%\begin{align}
n_{10} - n_{10} \cc = 0 & \Rightarrow c_1 (S) \cdot c_1 (\La) =  c_1 (S) \cdot c_1 (\Lc) \\
n_{10} + n_{10} \cc = 0 & \Rightarrow c_1 (\La)^2 + c_1 (\Lc)^2= -2
\end{align}

\newpage

\section{\label{B} F and D flatness conditions}

In the language of Table \ref{1}, the singlets $X_5$ and $\ov{X}_5$ correspond to $\theta_{45}$ and $\theta_{54}$ respectfully.  As these singlets get GUT scale VEVs in the E6SSM model, we must check that this is compatible with the $F$- and $D$-flatness conditions.  The D-flatness condition for $U_{A}(1)$ is

\begin{align}
\sum Q^A_{ij} (\left|\left\langle \theta_{ij} \right\rangle \right|^2 -\left|\left\langle \theta_{ji} \right\rangle \right|^2 ) & = - \frac{Tr Q^A}{192 \pi^2} g_s^2 M_S^2 \notag \\
& = - X Tr Q^A
\label{DflatE6}
\end{align}

\noindent This condition must be checked for all the U(1)s, the charge generators of which are given in the form $Q = {\rm diag}[t_1, t_2, t_3, t_4, t_5]$ by

\bea
Q_{\chi}&\propto &{\rm diag}[-1,-1,-1,-1,4]\\
Q_{\psi}&\propto &{\rm diag}[1,1,1,-3,0]\\
Q_{\perp}&\propto &{\rm diag}[1,1,-2,0,0]
\label{charges}
\eea

\noindent We can see immediately that if $\left\langle \theta_{45} \right\rangle = \left\langle \theta_{54} \right\rangle = M_{GUT}$, the presence of these VEVs will not affect the D flatness relations due to the relative minus sign in Eq. (\ref{DflatE6}).  As such, it is only necessary to check the conditions for F flatness.  As in the E6SSM model $\theta_{31}$ and $\theta_{53}$ get large VEVs while $\theta_{34}$ gets a TeV scale VEV, the only new problematic terms in the superpotential are

\be
W_{\theta} = \lambda_{ijk} \theta_{45}^i \theta_{53}^j \theta_{34}^k + M_{ab} \theta_{45}^a \theta_{54}^b
\ee

\noindent As such, the F flatness equations will be satisfied provided the following conditions are satisfied 

\begin{align*}
\frac{\partial W_{\theta}}{\partial \theta_{34}^k} &= \lambda_{ijk} \left\langle \theta_{45}^i \right\rangle \left\langle \theta_{53}^j \right\rangle = 0 \\
\frac{\partial W_{\theta}}{\partial \theta_{53}^j} &= \lambda_{ijk} \left\langle \theta_{45}^i \right\rangle \left\langle \theta_{34}^k \right\rangle = 0 \\
\frac{\partial W_{\theta}}{\partial \theta_{45}^i} &= \lambda_{ijk} \left\langle \theta_{53}^j \right\rangle \left\langle \theta_{34}^k \right\rangle + M_{ib} \left\langle \theta_{54}^b \right\rangle= 0 
\end{align*}

\noindent Due to the model building freedom we have in the number of singlet fields and the fact that the number of $\theta_{45}$ and $\theta_{54}$ fields in the spectrum can be changed by looking at topological relations where $\gamma > 4$ in Eq. (\ref{gamma}), these F flatness relations can always be satisfied in realisations of the E6SSM-like model.

\newpage

\section{\label{C} Anomaly Cancellation}

It has been noted in \cite{Palti:2012dd} that in models with multiple perpendicular U(1) symmetries, there is a $U(1)_{Y}-U(1)_{\perp}-U(1)_{\perp}$ anomaly which is not automatically cancelled through the spectral cover approach.  In order for this anomaly to be cancelled, the following condition is required:

\begin{equation}
3 \sum_{C_{10}^{i}} (Q_{10}^{i})^{A} (Q_{10}^{i})^{B} N_{10}^{i} + \sum_{C_{5}^{j}} (Q_{5}^{j})^{A} (Q_{5}^{j})^{B} N_{5}^{j} = 0 \label{anomaly}
\end{equation}

\noindent where the sums are over all the 10 and 5 matter curves, Q denotes the charge under either the U(1) labelled A or the one labelled B (allowing for mixed anomalies in the case of multiple U(1)s), and the Ns refer to the chirality induced by hypercharge flux.
In the models considered in this paper, we have 3 U(1)s, with generators

\begin{align}
Q_{\chi} &= \frac{1}{2\sqrt{10}}{\rm diag}(-1,-1,-1,-1,4) \\
Q_{\psi} &= \frac{1}{2\sqrt{6}}{\rm diag}(1,1,1,-3,0) \\
Q_{\perp} &= \frac{1}{2\sqrt{3}}{\rm diag}(1,1,-2,0,0)
\end{align}

\noindent As such, we can tabulate the U(1) charges of all the 5 and 10 matter curves in model 2.  The bulk exotics are not included in this table, as anomalies are automatically cancelled by the bulk spectrum.  As they come from a 78 of $E_6$, they are uncharged under $U(1)_{\perp}$, and so anomalies involving this U(1) are zero.  Also, as they always occur in vector pairs, the $U(1)_{\chi}$ and $U(1)_{\psi}$ anomalies are also cancelled.  Without the bulk states, the charges are shown in Table \ref{U(1) charges}.

\begin{table}[htdp]
\small
\centering
\begin{tabular}{|c|c|c|c|c|}
\hline
Curve& $Q_{\chi}$ & $Q_{\psi}$ & $Q_{\perp}$ & $N_Y$ \\
\hline
$10_{M}$ & $-\frac{1}{2 \sqrt{10}}$ & $\frac{1}{2 \sqrt{6}}$ & $\frac{1}{2 \sqrt{3}}$ & -1\\
\hline
$10_{2}$ & $-\frac{1}{2 \sqrt{10}}$ & $\frac{1}{2 \sqrt{6}}$ & $-\frac{1}{\sqrt{3}}$ & 1\\
\hline
$5_{H_u}$ & $\frac{1}{\sqrt{10}}$ & $-\frac{1}{\sqrt{6}}$ & $-\frac{1}{\sqrt{3}}$ & 1\\
\hline
$5_{1}$ & $\frac{1}{\sqrt{10}}$ & $-\frac{1}{\sqrt{6}}$ & $\frac{1}{2 \sqrt{3}}$ & -1\\
\hline
$5_{2}$ & $\frac{1}{\sqrt{10}}$ & $\frac{1}{\sqrt{6}}$ & $-\frac{1}{2 \sqrt{3}}$ & -1\\
\hline
$5_{3}$ & $-\frac{3}{2 \sqrt{10}}$ & $-\frac{1}{2 \sqrt{6}}$ & $-\frac{1}{2 \sqrt{3}}$ & -1\\
\hline
$5_{4}$ & $\frac{1}{\sqrt{10}}$ & $\frac{1}{\sqrt{6}}$ & $\frac{1}{\sqrt{3}}$ & 1\\
\hline
$5_{5}$ & $-\frac{3}{2 \sqrt{10}}$ & $-\frac{1}{2 \sqrt{6}}$ & $\frac{1}{\sqrt{3}}$ & 1\\
\hline
\end{tabular}
\caption{\small U(1) charges of the 10 and 5 matter curves}
\label{U(1) charges}
\end{table}%

\noindent We can now check if Eq. (\ref{anomaly}) holds for all the combinations of ${A,B}= {\chi, \psi, \perp}$ in $U(1)_{Y}-U(1)^{A}-U(1)^{B}$.  Plugging in the charges and the $N_Y$ values from Table \ref{U(1) charges} into the left hand side of Eq. (\ref{anomaly}) gives 

\begin{align*}
&A= \chi , B= \chi  \rightarrow 
3[- \frac{1}{40} + \frac{1}{40}]+[\frac{1}{10} - \frac{1}{10} - \frac{1}{10} - \frac{9}{40} + \frac{1}{10} + \frac{9}{40}] =0 \\
&A= \psi , B= \psi  \rightarrow 
3[-\frac{1}{24} +\frac{1}{24}] + [\frac{1}{6}- \frac{1}{6} - \frac{1}{6} - \frac{1}{24} + \frac{1}{6} + \frac{1}{24}] = 0 \\
&A= \chi , B= \psi  \rightarrow 
3[\frac{1}{4 \sqrt{60}} - \frac{1}{4 \sqrt{60}}]+[-\frac{1}{\sqrt{60}}+\frac{1}{\sqrt{60}}-\frac{1}{\sqrt{60}}-\frac{3}{4 \sqrt{60}} + \frac{1}{\sqrt{60}} + \frac{3}{4 \sqrt{60}}] = 0
\end{align*}

\noindent This shows that the relation  is indeed obeyed for the cases $U(1)_{Y}-U(1)^{\chi}-U(1)^{\chi}$, $U(1)_{Y}-U(1)^{\psi}-U(1)^{\psi}$ and $U(1)_{Y}-U(1)^{\chi}-U(1)^{\psi}$.  (This was to be expected, as $U(1)_{\chi}$ and $U(1)_{\psi}$ are both embedded in E6).  However, for the 3 anomalies involving $U(1)_{\perp}$, Eq. (\ref{anomaly}) is not satisfied, meaning that the anomalies involving $U(1)_{\perp}$ are not cancelled.  This $U(1)_{\perp}$, however, is broken by the $\theta_{31}$ VEV, so there are no problems below this scale.  Also, we know that all anomalies are automatically cancelled above the GUT scale so there is only a problem in the gap in energy between the GUT scale and $\theta_{31}$ VEV.  As we have $\left\langle \theta_{31}\right\rangle \approx 1.5 \times 10^{16} GeV$ and $M_{GUT} \approx 2 \times 10^{16} GeV$, the ratio of the GUT scale to the $U(1)_{\perp}$ breaking scale is only a factor of 1.5 and we do not regard anomalies in such a small energy interval as being problematic, especially bearing in mind the error in our energy scale estimates.

\end{appendix}

\end{document}